\documentclass[a4paper,10pt]{article}

\newcommand{\IEEEauthorblockN}[1]{#1}
\usepackage[margin=1.3in]{geometry}

\usepackage[utf8]{inputenc}
\usepackage[T1]{fontenc}
\usepackage{listings}
\usepackage[cmex10]{amsmath}
\usepackage{amsthm}
\usepackage{stmaryrd}
\usepackage{todonotes}
\usepackage{algorithm}
\usepackage[noend]{algorithmic}
\usepackage{url}
\usepackage{booktabs}
\usepackage[sort,compress]{cite}
\usepackage{tikz}
\usepackage[hidelinks,colorlinks=true,urlcolor=blue,citecolor=blue,linkcolor=blue]{hyperref}
\usetikzlibrary{decorations.markings}

\tikzstyle{cfgnode} = [draw, rectangle, minimum width=2cm, node distance=11mm]
\tikzstyle{arrowthicktip} = [decoration={markings,
    mark=at position 1 with {\arrow[scale=2]{>}}},
    postaction={decorate}]

\theoremstyle{definition}
\newtheorem{definition}{Definition}

\newcommand{\parag}[1]{\*\\{\bf #1}}
\newcommand{\rw}{\rightarrow}
\newcommand{\mmeth}[1]{\texttt{{#1}}}
\newcommand{\dsb}[1]{\llbracket \text{#1} \rrbracket}

\begin{document}

\title{Preventing Atomicity Violations with Contracts}

\author{
\IEEEauthorblockN{Diogo G. Sousa}
\and
\IEEEauthorblockN{Ricardo J.\ Dias}
\and
\IEEEauthorblockN{Carla Ferreira}
\and
\IEEEauthorblockN{João M.\ Lourenço}
}

\newcommand{\institute}{NOVA LINCS — NOVA Laboratory for Computer Science and Informatics\\
Departamento de Informática, Faculdade de Ciências e Tecnologia\\
Universidade NOVA de Lisboa, Portugal}

\date{\institute\\[2ex]\today}

\maketitle

\begin{abstract}
% What's the problem
% Why is it interesting
% What's the solution
% What follows from the solution
%
% Sincronização insuficiente
Software developers are expected to protect concurrent accesses to shared
regions of memory with some mutual exclusion primitive that ensures atomicity
properties to a sequence of program statements.
This approach prevents data races but may fail to provide all necessary
correctness properties.%, potentially leaving atomicity violations unaddressed.
% Violações de atomicidade
The composition of correlated atomic operations without further synchronization 
may cause atomicity violations. Atomic violations may be avoided by 
grouping the correlated atomic regions in a single larger atomic scope.
% Modules
Concurrent programs are particularly prone to atomicity violations 
when they use services provided by third party
packages or modules, since the programmer may fail to identify which
services are correlated.
% Solução: Contractos
In this paper we propose to use \emph{contracts for concurrency}, 
where the developer of a module writes a set of contract terms that specify which
methods are correlated and must be executed in the same atomic scope.
These contracts are then used to verify the correctness of the main program
with respect to the usage of the module(s). %, hence
% reducing the program errors due to atomicity violations.
If a contract is well defined and complete, and the main program respects it,
then the program is
safe from atomicity violations with respect to that module.
We also propose a static analysis based methodology to verify contracts for concurrency that we applied to some real-world software packages. The bug we found in Tomcat 6.0 was immediately acknowledged and corrected by its development team.
\end{abstract}

\section{Introduction}
\label{sec:intro}
%
% Motivação
The encapsulation of a set of functionalities as services of a software module offers
strong advantages in software development, since it promotes the reuse of code and
ease maintenance efforts.
If a programmer is unacquainted with the implementation details of a particular
set of services, she may fail to identify correlations that exist across those services,
such as data and code dependencies, leading to an inappropriate usage.
This is particularly relevant in a concurrent setting, where it is hard to account
for all the possible interleavings between threads and the effects of these
interleaved calls to the module's internal state.

% motivação: protocolos
One of the requirements for the correct behavior of a module is to respect its
\emph{protocol}, which defines the legal sequences of invocations to its methods.
For instance, a module that offers an abstraction to deal with files
typically will demand that the programmer start by calling the method
\lstinline{open()}, followed by an arbitrary number of \lstinline{read()} or
\lstinline{write()} operations, and concluding with a call to \lstinline{close()}.
A program that does not follow this protocol is incorrect and should be fixed.
A way to enforce a program to conform to such well defined behaviors is to use the design
by contract methodology~\cite{Meyer1992}, and specifying contracts that regulate
the module usage protocol.
In this setting, the contract not only serves as useful documentation, but may
also be automatically verified, ensuring the client's program obeys the module's
protocol~\cite{Cheon2007,Hurlin2009}.

% motivacao atomicidade protocolo
The development of concurrent programs brings new challenges on how to define
the protocol of a module.
Not only it is important to respect the module's protocol, but is also
necessary to guarantee the atomic execution of sequences of calls that are
susceptible of causing atomicity violations.
These atomicity violations are possible, even when the individual methods in the module are
protected by some concurrency control mechanism.
Figure~\ref{list:atomviolationexample} shows part of a program that schedules
tasks.
The \lstinline{schedule()} method gets a task, verifies if it is ready to run,
and execute it if so.
This program contains a potential atomicity violation since the method may
execute a task that is not marked as ready.  This may happen when another
thread concurrently schedules the same task, despite the fact the methods
of \lstinline{Task} are atomic.
In this case the \lstinline{isReady()} and \lstinline{run()} methods should
be executed in the same atomic context to avoid atomicity violations.
Atomicity violations are one of the most common source of bugs in concurrent
programming~\cite{Lu2008} and are particularly susceptible to occur when
composing calls to a module, as the developer may not be aware of the
implementation and internal state of the module.

\begin{figure}[t]
  \centering
  \begin{minipage}{0.68\linewidth}
    \begin{lstlisting}[frame=trbl]
void schedule() {
    Task t=taskQueue.next();

    if (t.isReady())
        t.run();
}
    \end{lstlisting}
  \end{minipage}\hfill
  \caption{Program with an atomicity violation.}
  \label{list:atomviolationexample}
\end{figure}

In this paper we propose to extend module usage protocols with
a specification of the sequences of calls that should be executed atomically.
We will also present an efficient static analysis to verify these protocols.

The contributions of this paper can be summarized as:

\begin{enumerate}
\item A proposal of \emph{contracts for concurrency} addressing the issue of atomicity violations;
\item A static analysis methodology to extract the behavior of
  a program with respect to the sequence of calls it may execute;
\item A static analysis to verify if a program conforms to a module's
  contract, hence that the module's correlated services are correctly invoked
  in the scope of an atomic region.
\end{enumerate}

The remaining of this paper is organized as follows.
In Section~\ref{sec:contract} we provide a specification and the semantics for
the contract. Section~\ref{sec:method} contains the general methodology of the
analysis.
Section~\ref{sec:programpattern} presents the phase of the analysis that extracts
the behavior of the client program while Section~\ref{sec:verification} shows how to
verify a contract based on the extracted information.
Section~\ref{sec:validation} follows with the presentation and discussion of the
results of our experimental validation.
The related work is presented in Section~\ref{sec:relwork}, and we conclude with
the final remarks in Section~\ref{sec:conclusion}.

\section{Contract Specification}
\label{sec:contract}
The contract of a module must specify which sequences of calls of its non-private
methods must be executed atomically, as to avoid atomicity violations in the
module's client program.
In the spirit of the \emph{programming by contract} methodology, we assume the
definition of the contract, including the identification of the sequences of
methods that should be executed atomically is a responsibility of the module's
developer.
\\\*
\line(1,0){250}\vspace{-3mm}
\begin{definition}[Contract]
  \label{def:contract}
  The contract of a module with public methods $m_1, \cdots\!, m_n$ is of the
  form,

  \begin{equation*}
    \begin{aligned}
      1. & \; e_1 \\
      2. & \; e_2 \\
      & \vdots    \\
      k. & \; e_k.
    \end{aligned}
  \end{equation*}

\noindent where each clause~$i$ is described by $e_i$, a star-free regular expression over the
  alphabet $\{ m_1, \cdots\!, m_n \}$.
  Star-free regular expressions are regular expressions without the Kleene star,
  using only the alternative ($|$) and the (implicit) concatenation operators.
\end{definition}
\vspace{-5mm}\hspace{-3mm}\*\line(1,0){250}
\vspace{1.5mm}

Each sequence defined in $e_i$ must be executed atomically by the program
using the module, otherwise there is a violation of the contract.
The contract specifies a finite number of sequences of calls, since it is the
union of star-free languages. Therefore, it is possible to have the same expressivity
by explicitly enumerating all sequences of calls, i.e., without using the
alternative operator.
We chose to offer the alternative operator so the programmer can group similar
scenarios under the same clause.
Our verification analysis assumes the contract defines a finite number of
call sequences.

%% \td{podemos tentar vender o peixe de outra maneira: dizemos que o modulo nao tem qq
%% tipo de sincronizacao. argumentamos que a sincronizacao deve ser feita pelo cliente
%% pois varios sequencias de metodos devem ser executadas no mesmo scope atomico.
%% Desta maneira contracto define: i) sequencias chamadas omissas nao precisam de
%% sincronização (por exemplo para obter valores constantes ou assim); ii) sequencias de
%% tamanho unitario só precisam de sincronizacao ao nivel do metodo; iii) sequencias
%% maiores implicam um scope atomico que abrange varios metodos. Isto tb tem a vantagem
%% de nao necessitar de adicionar sincronizacao a um modulo (por exemplo por ele só
%% existir compilado)}

\parag{Example}
Consider the array implementation offered by \emph{Java}
standard library, \lstinline{java.util.ArrayList}.
For simplicity we will only consider the methods
\lstinline{add(obj)}, \lstinline{contains(obj)}, \lstinline{indexOf(obj)},
\lstinline{get(idx)}, \lstinline{set(idx, obj)}, \lstinline{remove(idx)}, and
\lstinline{size()}.

The following contract defines some of the clauses for this class.

\begin{equation*}
  \label{eq:speceg1}
  \begin{aligned}
  1. & \; \mmeth{contains} \; \mmeth{indexOf} \\
  2. & \; \mmeth{indexOf} \; (\mmeth{remove} \; | \; \mmeth{set} \; | \; \mmeth{get})\\
  3. & \; \mmeth{size} \; (\mmeth{remove} \;
                           | \; \mmeth{set} \;
                           | \; \mmeth{get}) \\
  4. & \; \mmeth{add} \; \mmeth{indexOf}. \\
  \end{aligned}
\end{equation*}

Clause~$1$ of  \lstinline{ArrayList}'s contract denotes the
execution of \lstinline{contains()} followed by \lstinline{indexOf()} should
be atomic, otherwise the client program may confirm the
existence of an object in the array, but fail to obtain its index due to a
concurrent modification.
Clause~$2$ represents a similar scenario where, in addition, the position of
the object is modified.
In clause~$3$ we deal with the common situation where the program verifies if a
given index is valid before accessing the array.
To make sure the size obtained by \lstinline{size()} is valid when
accessing the array we should execute these calls atomically.
Clause~$4$ represents scenarios where an object is added to the array and
then the program tries to obtain information about that object by querying the
array.

Another relevant clause is
$\mmeth{contains} \; \mmeth{indexOf} \; (\mmeth{set} \; | \; \mmeth{remove}),$
but the contract's semantic already enforces the atomicity of this clause as a
consequence of the composition of clauses~$1$~and~$2$, as they overlap in the
\lstinline{indexOf()} method.

\section{Methodology}
\label{sec:method}
The proposed analysis verifies statically if a client program complies with the
contract of a given module, as defined in Section~\ref{sec:contract}.
This is achieved by verifying that the threads launched by the program always execute
atomically the sequence of calls defined by the contract.
\\

This analysis has the following phases:

\begin{enumerate}
\item Determine the entry methods of each thread launched by the program.
\item Determine which of the program's methods are atomically executed.
  We say that a method is \emph{atomically executed} if it is
  atomic\footnote{An atomic method is a method that explicitly applies a
    concurrency control mechanism to enforce atomicity.}
  or if the method is always called by atomically executed methods.
\item Extract the behavior of each of the program's threads with respect to the
  usage of the module under analysis.
\item For each thread, verify that its usage of the module respects the contract
  as defined in Section~\ref{sec:contract}.
\end{enumerate}

In Section~\ref{sec:programpattern} we introduce the algorithm that extracts the
program's behavior with respect to the module's usage.
Section~\ref{sec:verification} defines the methodology that verifies whether the
extracted behavior complies to the contract.

\section{Extracting the Behavior of a Program}
\label{sec:programpattern}
The behavior of the program with respect to the module usage can be seen as the
individual behavior of any thread the program may launch.
The usage of a module by a thread $t$ of a program can be described by a language
$L$ over the alphabet $m_1, \cdots\!, \, m_n$, the public methods of the module.
A word $m_1 \cdots \, m_n \in L$ if some execution of $t$ may run the sequence of
calls $m_1, \cdots\!, m_n$ to the module.

To extract the usage of a module by a program, our analysis generates a context-free
grammar that represents the language $L$ of a thread~$t$ of the client program,
which is represented by its control~flow~graph~(CFG)~\cite{Allen1970}.
The CFG of the thread~$t$ represents every possible path the control flow
may take during its execution.
In other words, the analysis generates a grammar $G_t$ such that, if there is an
execution path of $t$ that runs the sequence of calls $m_1, \cdots\!, m_n$, then
$m_1 \cdots \, m_n \in \mathcal{L}(G_t)$.
(The language represented by a grammar $G$ is denoted by $\mathcal{L}(G)$.)

A context-free grammar is especially suitable to capture the structure of the CFG
since it easily captures the call relations between methods that cannot be captured
by a weaker class of languages such as regular languages.
The first example bellow will show how this is done.
Another advantage of using context-free grammars
(as opposed to another static analysis technique) is that we can use efficient
algorithms for parsing to explore the language it represents.

\*
\hspace{-6mm}\line(1,0){250}\vspace{-1.5mm}
\begin{definition}[Program's Thread Behavior Grammar]
  \label{def:ppgrammar}
  The grammar $G_t=(N,\Sigma,P,S)$ is build from the CFG of the client's program
  thread~$t$.

  We define,

  \begin{itemize}
  \item $N$, the set of non-terminals, as the set of nodes of the CFG.
    Additionally we add non-terminals that represent each method of the
    client's program (represented in calligraphic font);
  \item $\Sigma$, the set of terminals,
    as the set of identifiers of the public methods of the module under analysis
    (represented in bold);
  \item $P$, the set of productions, as described bellow, by
    rules~\ref{eq:grammar0}--\ref{eq:grammar4};
  \item $S$, the grammar initial symbol, as the non-terminal that represents
    the entry method of the thread~$t$.
  \end{itemize}

  For each method \lstinline{f()} that thread~$t$ may run we add to $P$ the
  productions respecting the rules~\ref{eq:grammar0}--\ref{eq:grammar4}.
  Method \lstinline{f()} is represented by $\mathcal{F}$.
  A CFG node is denoted by $\alpha : \llbracket v \rrbracket$, where $\alpha$ is
  the non-terminal that represents the node and $v$ its type. We distinguish the
  following types of nodes:
  \emph{entry}, the entry node of  method $\mathcal{F}$;
  \emph{mod.h()}, a call to  method \lstinline{h()} of the module \emph{mod} under
  analysis;
  \emph{g()}, a call to another method \lstinline{g()} of the client program;
  and \emph{return}, the return point of  method $\mathcal{F}$.
  The $succ : N \rightarrow \mathcal{P}(N)$ function is used to obtain the successors
  of a given CFG node.

  \vspace{-2mm}

  \begin{align}
    \text{if } \alpha : \dsb{entry}, & \quad \{ \mathcal{F} \rw \alpha \}
    \cup \{ \alpha \rw \beta \; | \; \beta \in succ(\alpha) \} \subset P
    \label{eq:grammar0}\\
    \text{if } \alpha : \dsb{mod.h()}, & \quad
    \{ \alpha \rw {\bf h } \, \beta \; | \; \beta \in succ(\alpha) \} \subset P
    \label{eq:grammar1}\\
    \text{if } \alpha : \dsb{g()}, &  \quad
    \{ \alpha \rw \mathcal{G} \, \beta \; | \; \beta \in succ(\alpha) \} \subset P \nonumber\\
    & \qquad \qquad \quad \text{where $\mathcal{G}$ represents \texttt{g()}}
    \label{eq:grammar2}\\
    \text{if } \alpha : \dsb{return}, & \quad \{ \alpha \rw \epsilon \} \subset P
    \label{eq:grammar3}\\
    \text{if } \alpha : \dsb{otherwise}, & \quad \{ \alpha \rw \beta \; | \;
    \beta \in succ(\alpha) \} \subset P
    \label{eq:grammar4}
  \end{align}

  No more productions belong to $P$.
\end{definition}
\vspace{-4mm}\hspace{-5mm}\*\line(1,0){250}
\vspace{1.5mm}

Rules~\ref{eq:grammar0}--\ref{eq:grammar4} capture the structure of the CFG
in the form of a context-free grammar.
Intuitively this grammar represents the flow control of the thread~$t$ of the
program, ignoring everything not related with the module's usage.
For example, if ${ \bf f } \; { \bf g } \in \mathcal{L}(G_t)$ then the thread~$t$
may invoke, method \lstinline{f()}, followed by \lstinline{g()}.

Rule~\ref{eq:grammar0} adds a production that relates the non-terminal
$\mathcal{F}$, representing  method \lstinline{f()}, to the entry node of the
CFG of \lstinline{f()}.
Calls to the module under analysis are recorded in the grammar by the
Rule~\ref{eq:grammar1}.
Rule~\ref{eq:grammar2} handles calls to another method \lstinline{g()}
of the client program (method \lstinline{g()} will have its non-terminal
$\mathcal{G}$ added by Rule~\ref{eq:grammar0}).
The return point of a method adds an $\epsilon$ production to the grammar
(Rule~\ref{eq:grammar3}).
All others types of CFG nodes are handled uniformly, preserving the CFG structure
by making them reducible to the successor non-terminals (Rule~\ref{eq:grammar4}).
Notice that only the client program code is analyzed.

The $G_t$ grammar may be ambiguous, i.e., offer several different derivations
to the same word. Each ambiguity in the parsing of a sequence of calls
$m_1 \cdots \, m_n \in \mathcal{L}(G_t)$ represents different contexts
where these calls can be executed by thread~$t$.
Therefore we need to allow such ambiguities so that the verification
of the contract can cover all the occurrences of the sequences of calls in the
client program.

The language $\mathcal{L}(G_t)$ contains every sequence of calls the program
may execute, i.e., it produces no false negatives. However $\mathcal{L}(G_t)$ may
contain sequences of calls the program does not execute
(for instance calls performed inside a block of code that is never executed),
which may lead to false positives.

\parag{Examples}
\begin{figure*}[t]
  \centering
  \begin{minipage}{0.175\linewidth}
    \centering
    \begin{lstlisting}[frame=trbl]
void f() {
    m.a();
    if (cond)
        g();
    m.b();
}

void g() {
    m.c();
    if (cond) {
        g();
        m.d();
        f();
    }
}
    \end{lstlisting}
  \end{minipage}
  \hspace{8mm}
  \begin{minipage}{.45\linewidth}
    \begin{minipage}{\linewidth}
      \hspace{6mm} \lstinline{f()} \hspace{26mm} \lstinline{g()}
    \end{minipage}
    \hspace{-1mm}

    \begin{minipage}{0.20\linewidth}
      \centering
      \begin{tikzpicture}[scale=1]
        %    \draw[help lines] (0,0) grid (8,8);

        % nodes
        \node [cfgnode, name=A, node distance=10mm] {entry};

        \node [cfgnode, name=B, below of=A] {m.a()};
        \node [cfgnode, name=C, below of=B] {cond};
        \node [cfgnode, name=D, below of=C] {g()};
        \node [cfgnode, name=E, below of=D] {m.b()};
        \node [cfgnode, name=F, below of=E] {return};

        % labels
        \node [above of=A, left of=A, node distance=5mm] {A};
        \node [above of=B, left of=B, node distance=5mm] {B};
        \node [above of=C, left of=C, node distance=5mm] {C};
        \node [above of=D, left of=D, node distance=5mm] {D};
        \node [above of=E, left of=E, node distance=5mm] {E};
        \node [above of=F, left of=F, node distance=5mm] {F};

        % edges
        \draw [arrowthicktip] (A) -- (B);
        \draw [arrowthicktip] (B) -- (C);
        \draw [arrowthicktip] (C) -- (D);
        \draw [arrowthicktip] (D) -- (E);
        \draw [arrowthicktip] (E) -- (F);

        \draw [arrowthicktip] (C) -- +(1.5,0) |- (E);
      \end{tikzpicture}
    \end{minipage}
    \hspace{20mm}
    \begin{minipage}{0.20\linewidth}
      \begin{tikzpicture}[scale=1]
        %    \draw[help lines] (0,0) grid (8,8);

        % nodes
        \node [cfgnode, name=G] {entry};

        \node [cfgnode, name=H, below of=G] {m.c()};
        \node [cfgnode, name=I, below of=H] {cond};
        \node [cfgnode, name=J, below of=I] {g()};
        \node [cfgnode, name=K, below of=J] {m.d()};
        \node [cfgnode, name=L, below of=K] {f()};
        \node [cfgnode, name=M, below of=L] {return};

        % labels
        \node [above of=G, left of=G, node distance=5mm] {G};
        \node [above of=H, left of=H, node distance=5mm] {H};
        \node [above of=I, left of=I, node distance=5mm] {I};
        \node [above of=J, left of=J, node distance=5mm] {J};
        \node [above of=K, left of=K, node distance=5mm] {K};
        \node [above of=L, left of=L, node distance=5mm] {L};
        \node [above of=M, left of=M, node distance=5mm] {M};

        % edges
        \draw [arrowthicktip] (G) -- (H);
        \draw [arrowthicktip] (H) -- (I);
        \draw [arrowthicktip] (I) -- (J);
        \draw [arrowthicktip] (J) -- (K);
        \draw [arrowthicktip] (K) -- (L);
        \draw [arrowthicktip] (L) -- (M);

        \draw [arrowthicktip] (I) -- +(1.5,0) |- (M);
      \end{tikzpicture}
    \end{minipage}
    \hspace{8mm}
  \end{minipage}
  \caption{Program with recursive calls using the module \lstinline{m} (left)
    and respective CFG (right).}
  \label{fig:ppprogexample}
\end{figure*}
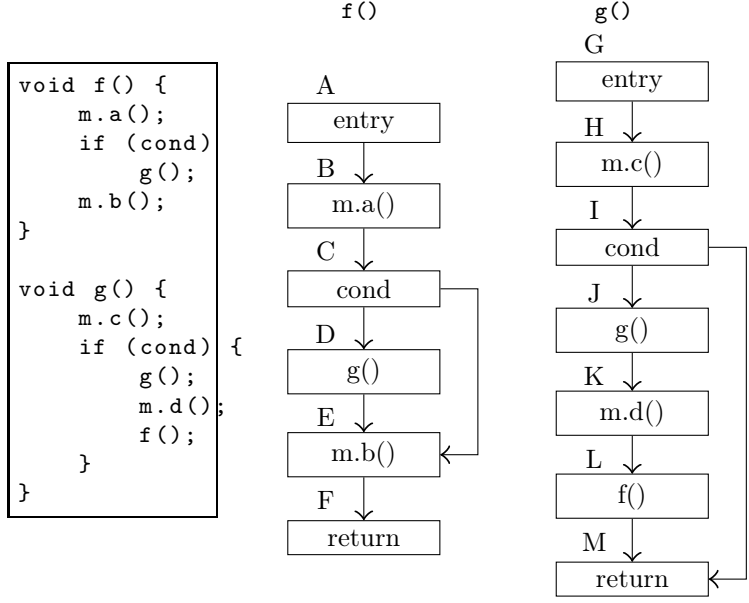
Figure~\ref{fig:ppprogexample}~(left) shows a program that consists of two
methods that call each other mutually.
Method~\lstinline{f()} is the entry point of the thread and
the module under analysis is represented by object~\lstinline{m}.
The control flow graphs of these methods are shown in
Figure~\ref{fig:ppprogexample}~(right).
According to Definition~\ref{def:ppgrammar}, we construct the grammar
$G_1=(N_1,\Sigma_1,P_1,S_1)$, where

\begin{align*}
  N_1      & = \{ \mathcal{F}, \mathcal{G}, A, B, C, D, E, F, G, H, I, J, K, L, M \}, \\
  \Sigma_1 & = \{ { \bf a, b, c, d } \}, \\
  S_1      & = \mathcal{F},
\end{align*}

and $P_1$ has the following productions:

\begin{equation*}
  \begin{aligned}
  \mathcal{F} & \rw A      \quad\qquad\qquad & \mathcal{G} & \rw G      \\
  A & \rw B                                  & H & \rw {\bf c } \, I    \\
  B & \rw {\bf a } \, C                      & I & \rw J \; | \; M      \\
  C & \rw D \; | \; E                        & J & \rw \mathcal{G} \, K \\
  D & \rw \mathcal{G} \, E                   & K & \rw {\bf d } \, L    \\
  E & \rw {\bf b } \, F                      & L & \rw \mathcal{F} \, M \\
  F & \rw \epsilon                           & M & \rw \epsilon.         \\
  \end{aligned}
\end{equation*}

%%%%%%%%%

\begin{figure*}[t]
  \centering
  \begin{minipage}{0.225\textwidth}
    \begin{lstlisting}[frame=trbl]
void f() {
    while (m.a()) {
        if (cond)
            m.b();
        else
            m.c();

        count++;
    }

    m.d();
}
    \end{lstlisting}
  \end{minipage}
  \hspace{10mm}
  \begin{minipage}{0.45\textwidth}
    \begin{tikzpicture}[scale=1]
      %    \draw[help lines] (0,0) grid (8,8);

      % nodes
      \node [cfgnode, name=A] {entry};

      \node [cfgnode, name=B, below of=A] {m.a()};
      \node [cfgnode, name=C, below of=B] {cond};

      \node [cfgnode, name=D, below of=C, left of=C] {m.b()};
      \node [cfgnode, name=E, below of=C, right of=C] {m.c()};

      \node [cfgnode, name=F, below of=C, node distance=22mm] {count++};

      \node [cfgnode, name=G, below of=F] {m.d()};
      \node [cfgnode, name=H, below of=G] {return};

      % labels
      \node [above of=A, left of=A, node distance=5mm] {A};
      \node [above of=B, left of=B, node distance=5mm] {B};
      \node [above of=C, left of=C, node distance=5mm] {C};
      \node [above of=D, left of=D, node distance=5mm] {D};
      \node [above of=E, right of=E, node distance=5mm] {E};
      \node [right of=F, node distance=13mm] {F};
      \node [above of=G, left of=G, node distance=5mm] {G};
      \node [above of=H, left of=H, node distance=5mm] {H};

      % edges
      \draw [arrowthicktip] (A) -- (B);
      \draw [arrowthicktip] (B) -- (C);
      \draw [arrowthicktip] (C) -- (D);
      \draw [arrowthicktip] (C) -- (E);
      \draw [arrowthicktip] (D) -- (F);
      \draw [arrowthicktip] (E) -- (F);

      \draw [arrowthicktip] (F) -- +(-2.5,0) |- (B);
      \draw [arrowthicktip] (B) -- +(2.5,0) |- (G);

      \draw [arrowthicktip] (G) -- (H);
    \end{tikzpicture}
  \end{minipage}
  \caption{Program using the module \lstinline{m} (left) and respective CFG (right).}
  \label{fig:ppprogexample2}
\end{figure*}
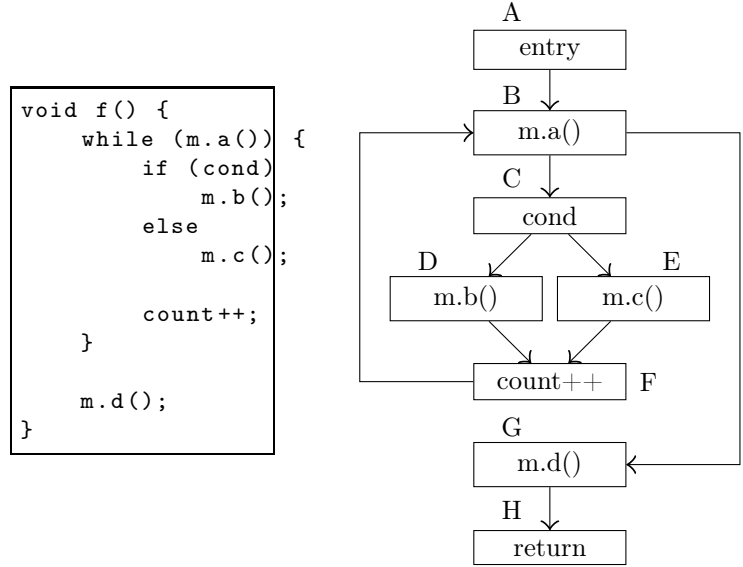

A second example, shown in Figure~\ref{fig:ppprogexample2}, exemplifies how
the Definition~\ref{def:ppgrammar} handles a flow control with loops.
In this example we have a single function \lstinline{f()}, which is assumed to be
the entry point of the thread. We have $G_2=(N_2,\Sigma_2,P_2,S_2)$, with
\begin{align*}
  N_2      & = \{ \mathcal{F}, A, B, C, D, E, F, G, H\}, \\
  \Sigma_2 & = \{ { \bf a, b, c, d }\}, \\
  S_2      & = \mathcal{F}.
\end{align*}

The set of productions $P_2$ is,
\begin{equation*}
  \begin{aligned}
    \mathcal{F} & \rw A        \quad\qquad\qquad & E & \rw { \bf c } \, F \\
    A & \rw B                                    & F & \rw B \\
    B & \rw { \bf a } \, C \; | \; { \bf a } \, G
                                                 & G & \rw { \bf d } \, H \\
    C & \rw D \; | \; E
                                                 & H & \rw \epsilon\\
    D & \rw { \bf b } \, F.
  \end{aligned}
\end{equation*}

\section{Contract Verification}
\label{sec:verification}
The verification of a contract must ensure all sequences of calls specified
by the contract are executed atomically by all threads the client
program may launch.
Since there is a finite number of call sequences defined by the contract we can
verify each of these sequences to check if the contract is respected.
\begin{algorithm}[t]
  \caption{Contract verification algorithm.}
  \label{algo:verification}
  \begin{algorithmic}[1]
    \REQUIRE{$P$, client's program;
      \\\hspace{9.95mm}$C$, module contract (set of allowed sequences).}
    \FOR{$t \in \text{threads}(P)$}
      \STATE{$G_t \gets \text{make\_grammar}(t)$}
      \STATE{$G_t' \gets \text{subword\_grammar}(G_t)$}
      \FOR{$w \in C$}
        \STATE{$T \gets \text{parse}(G_t',w)$}
        \FOR{$\tau \in T$}
          \STATE{$N \gets \text{lowest\_common\_ancestor}(\tau,w)$}
          \IF{$\neg \text{run\_atomically}(N)$}
            \RETURN{ERROR}
          \ENDIF
        \ENDFOR
      \ENDFOR
    \ENDFOR
    \RETURN{OK}
  \end{algorithmic}
\end{algorithm}

The idea of the algorithm is to generate a grammar the captures the behavior of
each thread with respect to the module usage.  Any sequence of the calls
contained in the contract can then be found by parsing the word
(i.e. the sequence of calls) in that grammar.  This will create a parsing tree
for each place where the thread can execute that sequence of calls.
The parsing tree can then be inspected to determine the atomicity of the sequence
of calls discovered.

Algorithm~\ref{algo:verification} presents the pseudo-code of the algorithm
that verifies a contract against a client's program.
For each thread~$t$ of a program~$P$, it is necessary to determine if (and where)
any of the sequences of calls defined by the contract $w = m_1, \cdots\!, m_n$
occur in $P$ (line 4).
To do so, each of the these sequences are parsed in the grammar $G_t'$ (line 5)
that includes all words and sub-words of $G_t$.
Sub-words must be included since we want to take into account partial traces of the
execution of thread~$t$, i.e., if we have a program
\lstinline{m.a(); m.b(); m.c(); m.d();} we are able to verify the word
${ \bf b } \; { \bf c }$ by parsing it in $G_t'$.
Notice that $G_t'$ may be ambiguous. Each different parsing tree represents different
locations where the sequence of calls $m_1, \cdots\!, m_n$ may occur in thread~$t$.
Function $\mmeth{parse()}$ returns the set of these parsing trees.
Each parsing tree contains information about the location of each methods call
of $m_1, \cdots\!, m_n$ in program $P$
(since non-terminals represent CFG nodes).
Additionally, by going upwards in the parsing tree, we can find the node that
represents the method under which all calls to $m_1, \cdots\!, m_n$ are performed.
This node is the lowest common ancestor of  terminals $m_1, \cdots\!, m_n$ in
the parsing tree (line 7). Therefore we have to check the lowest common ancestor
is always executed atomically (line 8) to make sure the whole sequence of calls is
executed under the same atomic context.
Since it is the \emph{lowest} common ancestor we are sure to require the
minimal synchronization from the program.
A parsing tree contains information about the location in the program
where a contract violation may occur, therefore we can offer detailed
instructions to the programmer on where this violation occurs and
how to fix it.

Grammar $G_t$ can use all the expressivity offered by context-free languages.
For this reason it is not sufficient to use the $LR(\cdot)$ parsing
algorithm~\cite{Knuth1965}, since it does not handle ambiguous grammars.
To deal with the full class of context-free languages a GLR parser
(Generalized $LR$ parser) must be used. GLR parsers explore all the ambiguities
that can generate different derivation trees for a word.
A GLR parser was introduced by Tomita in \cite{Tomita1987}. Tomita presents a
non-deterministic versions of the $LR(0)$ parsing algorithm with some optimizations
in the representation of the parsing stack that improve the temporal and spacial
complexity of the parsing phase.

Another important point is that the number of parsing trees may be infinite.
This is due to loops in the grammar, i.e., derivations from a non-terminal to itself
($A \Rightarrow \cdots \Rightarrow A$), which often occur in $G_t$
(every loop in the control flow graph will yield a corresponding loop in the grammar).
For this reason the parsing algorithm must detect and prune parsing branches
that will lead to redundant loops, ensuring a finite number of parsing trees is
returned.
To achieve this the parsing algorithm must detect a loop in the list of reduction
it has applied in the current parsing branch, and abort it if that loop did not
contribute to parse a new terminal.

\parag{Examples}
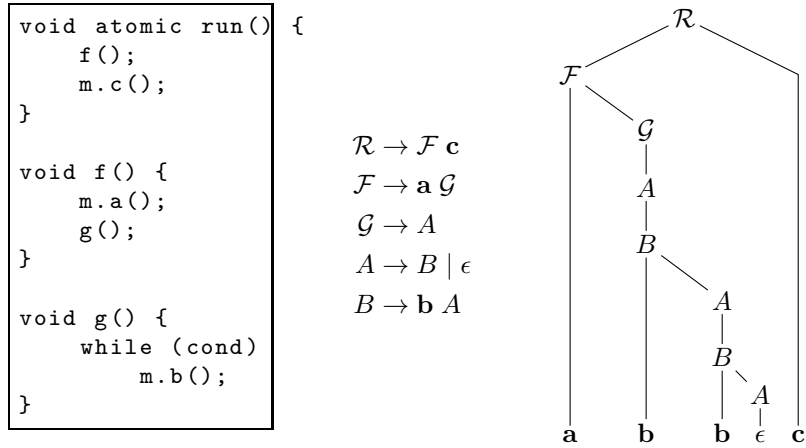
\begin{figure*}[t]
  \centering
  \begin{minipage}{0.225\linewidth}
  \begin{lstlisting}[frame=trbl]
void atomic run() {
    f();
    m.c();
}

void f() {
    m.a();
    g();
}

void g() {
    while (cond)
        m.b();
}
\end{lstlisting}
 \end{minipage}
  \begin{minipage}{0.25\linewidth}
    \begin{equation*}
      \begin{aligned}
        \mathcal{R} & \rw \mathcal{F} \; { \bf c } \\
        \mathcal{F} & \rw { \bf a } \; \mathcal{G} \\
        \mathcal{G} & \rw A \\
        A           & \rw B \; | \; \epsilon \\
        B           & \rw { \bf b } \; A \\
      \end{aligned}
    \end{equation*}
  \end{minipage}
  \begin{minipage}{0.25\linewidth}
    \begin{tikzpicture}[inner sep=.5mm]
      % \draw[help lines] (0,0) grid (7,8);

      \node[anchor=south] (a)  at (1, 0.5) {$ \bf a$};
      \node[anchor=south] (b1) at (2, 0.5) {$ \bf b$};
      \node[anchor=south] (b2) at (3, 0.5) {$ \bf b$};
      \node[anchor=south] (c)  at (4, 0.5) {$ \bf c$};

      \node[anchor=south] (B1) at (3, 1.5)  {$B$};
      \node[anchor=south] (A1) at (3, 2.25) {$A$};
      \node[anchor=south] (A2) at (3.5, 1)  {$A$};

      \node[anchor=south] (e) at (3.5, .5)  {$\epsilon$};

      \node[anchor=south] (G)  at (2, 4.5)  {$\mathcal{G}$};
      \node[anchor=south] (A3) at (2, 3.75)  {$A$};
      \node[anchor=south] (B2) at (2, 3)  {$B$};

      \node[anchor=south] (F)  at (1, 5.25)  {$\mathcal{F}$};

      \node[anchor=south] (R)  at (2.5, 6)  {$\mathcal{R}$};

      \path[-] (R) edge (F);
      \draw (R) -- +(1.5,-.75) -- (c);

      \path[-] (F) edge (a);
      \path[-] (F) edge (G);
      \path[-] (G) edge (A3);
      \path[-] (A3) edge (B2);
      \path[-] (B2) edge (A1);
      \path[-] (B2) edge (b1);
      \path[-] (A1) edge (B1);
      \path[-] (B1) edge (A2);
      \path[-] (B1) edge (b2);
      \path[-] (A2) edge (e);
    \end{tikzpicture}
  \end{minipage}
  \caption{Program (left), simplified grammar (center) and parsing tree of
    ${\bf a \; b \; b \; c }$ (right).}
  \label{fig:parse}
\end{figure*}
Figure~\ref{fig:parse} shows a program (left), that uses the module
\lstinline{m}. The method \lstinline{run()} is the entry point of the thread~$t$
and is atomic.
In the center of the figure we shown a simplified version of the $G_t$ grammar.
(The $G_t'$ grammar is not shown for the sake of brevity.)
The \lstinline{run()}, \lstinline{f()}, and \lstinline{g()} methods are
represented in the grammar by the non-terminals $\mathcal{R}$, $\mathcal{F}$,
and $\mathcal{G}$ respectively.
If we apply  Algorithm~\ref{algo:verification} to this program with the
contract $C=\{ { \bf a \; b \; b \; c } \}$ the resulting parsing tree,
denoted by $\tau$ (line $6$ of Algorithm~\ref{algo:verification}), is
represented in Figure~\ref{fig:parse} (right).
To verify all calls represented in this tree are executed atomically, the
algorithm determines the lowest common ancestor of
${ \bf a \; b \; b \; c }$ in the parsing tree (line $7$), in this
example $\mathcal{R}$.
Since $\mathcal{R}$ is always executed atomically (\lstinline{atomic} keyword),
it complies to the contract of the module.

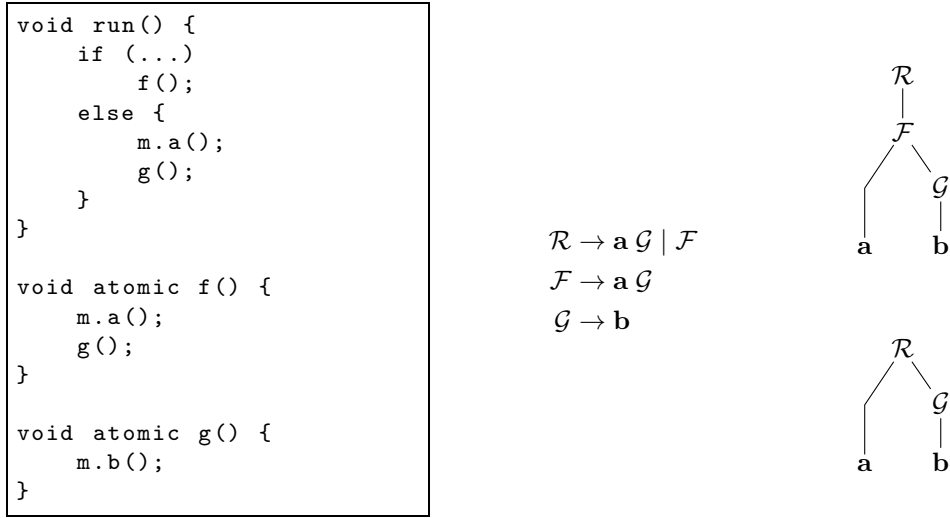
\begin{figure}[t]
  \centering
  \begin{minipage}{0.37\linewidth}
  \begin{lstlisting}[frame=trbl]
void run() {
    if (...)
        f();
    else {
        m.a();
        g();
    }
}

void atomic f() {
    m.a();
    g();
}

void atomic g() {
    m.b();
}
\end{lstlisting}
 \end{minipage}
  \begin{minipage}{0.35\linewidth}
    \centering
    \begin{equation*}
      \begin{aligned}
        \mathcal{R} & \rw { \bf a } \; \mathcal{G} \; | \; \mathcal{F} \\
        \mathcal{F} & \rw { \bf a } \; \mathcal{G} \\
        \mathcal{G} & \rw { \bf b }
      \end{aligned}
    \end{equation*}
  \end{minipage}
  \begin{minipage}{0.15\linewidth}
    \centering
    \begin{tikzpicture}[inner sep=.5mm]
      % \draw[help lines] (0,0) grid (7,8);

      \node[anchor=south] (a) at (1, 0) {$ \bf a$};
      \node[anchor=south] (b) at  (2, 0) {$ \bf b$};

      \node[anchor=south] (G)  at (2, .75)  {$\mathcal{G}$};
      \node[anchor=south] (R)  at (1.5, 2.25)  {$\mathcal{R}$};

      \node[anchor=south] (F)  at (1.5, 1.5)  {$\mathcal{F}$};

      \path[-] (R) edge (F);
      \path[-] (F) edge (G);
      \path[-] (G) edge (b);
      \draw (F) -- +(-.5,-.75) -- (a);
    \end{tikzpicture}

    \vspace{10mm}

    \begin{tikzpicture}[inner sep=.5mm]
      % \draw[help lines] (0,0) grid (7,8);

      \node[anchor=south] (a) at (1, 0) {$ \bf a$};
      \node[anchor=south] (b) at  (2, 0) {$ \bf b$};

      \node[anchor=south] (G)  at (2, .75)  {$\mathcal{G}$};

      \node[anchor=south] (R)  at (1.5, 1.5)  {$\mathcal{R}$};

      \path[-] (R) edge (G);
      \path[-] (G) edge (b);
      \draw (R) -- +(-.5,-.75) -- (a);
    \end{tikzpicture}
  \end{minipage}
  \caption{Program (left), simplified grammar (center) and parsing tree of
    ${\bf a \; b }$ (right).}
  \label{fig:parse2}
\end{figure}

Figure~\ref{fig:parse2} exemplifies a situation where the generated grammar is
ambiguous. In this case the contract is $C = \{ { \bf a \; b } \}$.
The figure shows the two distinct ways to parse the word
${\bf a \; b}$ (right).
Both these trees will be obtained by our verification algorithm
(line 5 of Algorithm~\ref{algo:verification}).
The first tree (top) has $\mathcal{F}$ as the lowest common ancestor of
${\bf a \; b}$. Since $\mathcal{F}$ corresponds to the method \lstinline{f()},
which is executed atomically, so this tree respects the contract.
The second tree (bottom) has $\mathcal{R}$ as the lowest common ancestor of
${\bf a \; b }$, corresponding to the execution of the \lstinline{else} branch
of method \lstinline{run()}.
This non-terminal ($\mathcal{R}$) does not correspond to an atomically executed method,
therefore the contract is not met and a contract violation is detected.

\section{Analysis with Points-to}
\label{sec:pointsto}
In an object-oriented programming language the module is often represented as an
object, in which case we should differentiate the instances of the class of the
module. This section explains how the analysis is extended to handle multiple
instances of the module by using \emph{points-to} information.

\begin{algorithm}[t]
  \caption{Contract verification algorithm with points-to information.}
  \label{algo:verificationpointsto}
  \begin{algorithmic}[1]
    \REQUIRE{$P$, client's program;
      \\\hspace{9.95mm}$C$, module contract (set of allowed sequences).}
    \FOR{$t \in \text{threads}(P)$}
      \FOR{$a \in \text{mod\_alloc\_sites}(t)$}
        \STATE{$G_{t_a} \gets \text{make\_grammar}(t,a)$}
        \STATE{$G_{t_a}' \gets \text{subword\_grammar}(G_{t_a})$}
        \FOR{$w \in C$}
          \STATE{$T \gets \text{parse}(G_{t_a}',w)$}
          \FOR{$\tau \in T$}
            \STATE{$N \gets \text{lowest\_common\_ancestor}(\tau,w)$}
            \IF{$\neg \text{run\_atomically}(N)$}
              \RETURN{ERROR}
            \ENDIF
          \ENDFOR
        \ENDFOR
      \ENDFOR
    \ENDFOR
    \RETURN{OK}
  \end{algorithmic}
\end{algorithm}

To extend the analysis to points-to a different grammar is generated for each
allocation site of the module.  Each allocation site represents an instance of
the module, and the verification algorithm verifies the contract words for
each allocation site and thread (whereas the previous algorithm verified the
contract words for each thread).
The new algorithm is shown in Algorithm~\ref{algo:verificationpointsto}.
It generated the grammar $G_{t_a}$ for a thread $t$ and module instance $a$.
This grammar can be seen as the behavior of thread $t$ with respect to
the module instance $a$, ignoring every other instance of the module.

To generate the grammar $G_{t_a}$ we adapt the Definition~\ref{def:ppgrammar}
to only take into account the instance $a$.
The grammar generation is extended in the following way:

\*
\hspace{-6mm}\line(1,0){250}\vspace{-1.5mm}
\begin{definition}[Program's Thread Behavior Grammar with points-to]
  \label{def:ppgrammarpointsto}
  The grammar $G_t=(N,\Sigma,P,S)$ is build from the CFG of the client's program
  thread~$t$ and an object allocation site $a$, which represents an instance of the
  module.

  We define $N$, $\Sigma$, $P$ and $S$ in the same way as
  Definition~\ref{def:ppgrammar}.

  The rules remain the same, except for rule~\ref{eq:grammar1}, which becomes:

  \begin{align}
    & \text{if } \alpha : \dsb{mod.h()} \text{ and mod can only point to } a
        \label{eq:grammarpt0}\\
    & \quad
    \{ \alpha \rw {\bf h } \, \beta \; | \; \beta \in succ(\alpha) \} \subset P
    \nonumber\\
    & \text{if } \alpha : \dsb{mod.h()} \text{ and mod can point to } a
    \label{eq:grammarpt1}\\
    & \quad
    \{ \alpha \rw {\bf h } \, \beta \; | \; \beta \in succ(\alpha) \} \subset P
    \nonumber\\
    & \quad
    \{ \alpha \rw \beta \; | \; \beta \in succ(\alpha) \} \subset P
    \nonumber\\
    & \text{if } \alpha : \dsb{mod.h()} \text{ and mod cannot point to } a
        \label{eq:grammarpt2}\\
    & \quad
    \{ \alpha \rw \beta \; | \; \beta \in succ(\alpha) \} \subset P     \nonumber
  \end{align}
\end{definition}
\vspace{-4mm}\hspace{-5mm}\*\line(1,0){250}
\vspace{1.5mm}

Here we use the points-to information to generate the grammar, and we should
consider the places where a variable can point-to.
If it may point-to our instance $a$ or another instance we consider both
possibilities in the Rule~\ref{eq:grammarpt1} of
Definition~\ref{def:ppgrammarpointsto}.

\section{Contracts with Parameters}
\label{sec:contractwithargs}
\begin{figure}[t]
  \begin{lstlisting}
void replace(int o, int n)
{
    if (array.contains(o))
    {
        int idx=array.indexOf(o);
        array.set(idx,n);
    }
}
  \end{lstlisting}

  \caption{Examples of atomic violation with data dependencies.}
  \label{list:atomviolationdf}
\end{figure}
Frequently contract clauses can be refined by considering the flow of data
across calls to the module. For instance Listing~\ref{list:atomviolationdf}
shows a procedure that replaces an item in an array by another.
This listing contains two atomicity violations: the element might not exist
when \lstinline{indexOf()} is called; and the index obtained might be outdated
when \lstinline{set()} is executed.
Naturally, we can define a clause that forces the atomicity of this sequence
of calls as $\mmeth{contains} \; \mmeth{indexOf} \; \mmeth{set}$, but this can
be substantially refined by explicitly require that a correlation exists
between the \lstinline{indexOf()} and \lstinline{set()} calls.
To do so we extend the contract specification to capture the arguments and
return values of the calls, which allows the user to establish the relation of
values across calls.

The contract can therefore be extended to accommodate this relations,
in this case the clause might be
\begin{equation*}
  \mmeth{contains(X)} \;\;\; \mmeth{Y=indexOf(X)} \;\;\; \mmeth{set(Y,\_)}.
\end{equation*}

This clause contains variables ($\mmeth{X}, \mmeth{Y}$) that must
satisfy unification for the clause to be applicable. The underscore
symbol ($\mmeth{\_}$) represents a variable that will not be used
(and therefore requires no binding).
Algorithm~\ref{algo:verification} can easily be modified to filter out the
parsing trees that correspond to calls that do not satisfy the unification
required by the clause in question.

In our implementation we require a exact match between the terms of the program
to satisfy the unification, since it was sufficient for most scenarios.
It can however be advantageous to generalize the unification relation.
For example, the calls
\begin{lstlisting}
    array.contains(o);
    idx=array.indexOf(o+1);
    array.set(idx,n);
\end{lstlisting}

also imply a data dependency between the first two calls.
We should say that $A$ unifies with $B$ if, and only if, the value of $A$
depends on the value of $B$, which can occur due to value manipulation
(data dependency) or control-flow dependency (control dependency).
This can be obtained by an information flow analysis, such as presented
in~\cite{Bergeretti1985}, which can statically infer the variables that
influenced the value that a variable hold on a specific part of the program.

This extension of the analysis can be a great advantage
for some types of modules. As an example we rewrite the contract
for the \emph{Java} standard library class, \lstinline{java.util.ArrayList},
presented in Section~\ref{sec:contract}:

\begin{equation*}
  \label{eq:speceg2}
  \begin{aligned}
  1. & \; \mmeth{contains(X)} \; \mmeth{indexOf(X)} \\
  2. & \; \mmeth{X=indexOf(\_)} \; (\mmeth{remove(X)} \;
                                   | \; \mmeth{set(X,\_)} \;
                                   | \; \mmeth{get(X)})\\
  3. & \; \mmeth{X=size()} \; (\mmeth{remove(X)} \;
                             | \; \mmeth{set(X,\_)} \;
                             | \; \mmeth{get(X)}) \\
  4. & \; \mmeth{add(X)} \; \mmeth{indexOf(X)}.
  \end{aligned}
\end{equation*}

This contract captures in detail the relations between calls that may be
problematic, and excludes from the contract sequences of calls that does not
constitute atomicity violations.

\section{Prototype}
\label{sec:prototype}
%
% prototype setting
A prototype was implemented to evaluate our methodology. This tool analyses
\emph{Java} programs using Soot~\cite{Vallee-Rai1999}, a Java static analysis
framework.
This framework directly analyses \emph{Java bytecode}, allowing us to analyse a
compiled program, without requiring access to its source code.
In our implementation a method can be marked atomic with a \emph{Java} annotation.
The contract is also defined as an annotation of the class representing the
module under analysis. The prototype is available in
\url{https://github.com/trxsys/gluon}.

\subsection{Optimizations}
To achieve a reasonable time performance we implemented a few optimizations.
Some of these optimizations reduced the analysis run time by a few orders of
magnitude in some cases, without sacrificing precision.

A simple optimization was applied to the grammar to reduce its size.
When constructing the grammar, most control flow graph nodes will have a single
successor.
Rule~\ref{eq:grammar4} (Definition~\ref{def:ppgrammar}) will always be
applied to these kind of nodes, since they represent an instruction
that does not call any function.  This creates redundant ambiguities in the grammar
due to the multiple control flow paths that never use the module under analysis.
To avoid exploration of redundant parsing
branches we rewrite the grammar to transform productions of the form
$A \rw \beta B \delta, \, B \rw \alpha$ to $A \rw \beta \alpha \delta$, if
no other rule with head $B$ exists.
For example, an \lstinline{if else} that do not use the module will create
the productions $A \rw B, \, A \rw C, \, B \rw D, \, C \rw D$.
This transformation will reduce it to $A \rw D$, leaving no ambiguity for the
parser to explore here.
This optimization reduced the analysis time by at least one order of magnitude
considering the majority of the tests we performed.
For instance, the Elevator test could not be analyzed in a reasonable time prior
to this optimization.
%The cost of this grammar reduction is negligible: it was performed in less than
%$6 \mathrm{s}$ for all presented tests.

Another optimization was applied during the parsing phase. Since the $GLR$
parser builds the derivation tree bottom-up we can be sure to find the lowest
common ancestor of the terminals as early as possible. The lowest common ancestors
will be the first non-terminal in the tree covering all the terminals of the
parse tree.
This is easily determined if we propagate, bottom-up, the number of terminals
each node of the tree covers.
Whenever a lowest common ancestor is determined we do not need further parsing
and can immediately verify if the corresponding calls are in the same atomic
context.  This avoids completing the rest of the tree which can contain
ambiguities, therefore a possibly large number of new branches is avoided.

Another key aspect of the parsing algorithm implementation is the loop detection.
To achieve a good performance we should prune parsing branches that generated
unproductive loop as soon as possible. Our implementation guarantees
the same non-terminal never appears twice in a parsing tree without contributing
to the recognition of a new terminal.

To achieve a better performance we also do not explicitly compute the subword
grammar ($G_t'$).  We have modified our $GLR$ parser to parse subwords as
described in~\cite{rekers1991}. This greatly improves the parser performance
because constructing $G_t'$ introduces many irrelevant ambiguities the parser
had to explore.

\begin{table*}[t]
  \centering
  \caption{Optimization Improvements.}
  \label{tab:opt}
  \begin{tabular}{l c@{\hspace{6mm}}}
    \toprule
        {\bf Optimization} & {\bf Improvement} \\
    \midrule
    Grammar Simplification     &   428\% \\
    Stop Parsing at LCA        &   ? \\
    Subword Parser             &   3\% \\
    \bottomrule
  \end{tabular}
\end{table*}

The Table~\ref{tab:opt} show how much the of the optimizations improve
the analysis performance.  These results are build from an test made to
stress the performance of gluon but is consistent with real applications.
The \emph{Improvement} column show how much of an improvement that particular
optimization contributes to the analysis.
The \emph{Stop Parsing at LCA} cause an improvement that we were not able to
measure since the test was unable to complete in reasonable time.

% impact of disabling each of these optimizations:
%
% grammar simpl. 305.930s
% stop at LCA    infeasible
% subword parser 74.182s
%
% normal run     71.467s
%
% with test ''performance'' in tests/simple/performance.

\subsection{Class Scope Mode}
\label{sec:classscopemode}
Gluon normally analyzes the entire program, taking into account any sequence of
calls that can spread across the whole program (as long as they are consecutive
calls to a module).  However this is infeasible for very large programs so, for
these programs, we ran the analysis with for each class, ignoring calls to other
classes.  This will detect contract violations where the control flow does not
escape the class, which is reasonable since code locality indicates a stronger
correlations between calls.

This mode of operation can be useful to analyze large programs as they might
have very complex control flow graphs and thus are infeasible to analyze with
the scope of the whole program.

In this mode the grammar is build for each class instead of each thread.
The methods of the class will create non-terminals
$\mathcal{F}_1, \cdots\!, \mathcal{F}_n$, just as before.
The only change in creating this grammar is that we create the productions
$S \rw \mathcal{F}_1 \; | \; \cdots \; | \; \mathcal{F}_n$ as the starting
production of the grammar ($S$ being the initial symbol).
This means that we consider the execution of all methods of the class being
analyzed.

\section{Validation and Evaluation}
\label{sec:validation}
% validation methodology
To validate the proposed analysis we analyzed a few real-world programs
(Tomcat, Lucene, Derby, OpenJMS and Cassandra) as well as small programs
known to contain atomicity violations.  These small programs were adapted
from the literature~\cite{Praun2004,Artho03,Artho04,Teixeira2011,IBM,PessanhaMSc,
Beckman2008} and are typically used to evaluate atomicity violation detection
techniques.
We modified these small programs to employ a modular design and we wrote
contracts to enforce the correct atomic scope of calls to that module.
Some additional clauses were added that may represent atomicity violations
in the context of the module usage, even if the program do not violate those
clauses.

For the large benchmarks analyzed we aimed to discover new, unknown, atomicity
violations.  To do so we had to create contracts in an automated manner, since
the code base was too large.
To automate the generation of contracts we employ a very simplistic approach that
tries to infer the contract's clauses based on what is already synchronized in the
code.  This idea is that most sequences of calls that should be atomic was
correctly used \emph{somewhere}.  Having this in mind we look for sequences
of calls done to a module that are used atomically at least two points of the
program.  If a sequence of calls is done atomically in two places of the code
that might indicate that these calls are correlated and should be atomic.
We used these sequences as our contracts, after manually filtering a few
irrelevant contracts.
This is a very simple way to generate contracts, which should ideally be written
by the module's developer to capture common cases of atomicity violations, so we
can expect the contracts to be more fine-tuned to better target atomicity violations
if the contracts are part of the regular project development.

Since these programs load classes dynamically it is impossible to obtain complete
points-to information, so we are pessimistic
and assumed every module instance could be referenced by any variable that are
type-compatible.
We also used the class scope mode described in Section~\ref{sec:classscopemode}
because it would be impractical to analyze such large programs with the scope of
the whole program.
This restrictions did not apply to the small programs analyzed.

% #clauses
% #violations
% false positives
%  convervative points-to
% potencial atomicity violations (if it was shared)
% true atomicity violations
% SLOC
% Time

\begin{table*}[t]
  \centering
  \caption{Validation results.}
  \label{tab:valresults}
  \resizebox{\textwidth}{!}{
  \begin{tabular}{@{} l c@{\hspace{6mm}} c@{\hspace{6mm}}  c@{\hspace{6mm}}
      c@{\hspace{6mm}} c@{\hspace{6mm}} c@{\hspace{6mm}} c@{\hspace{6mm}}
      | c@{\hspace{6mm}} c@{}}
    \toprule
            &  & {\bf Contract}
      & {\bf False} & {\bf Potential} & {\bf Real} & 
      & & {\bf ICFinder} & {\bf ICFinder} \\
%        & & & {\bf \scriptsize (Conservative points-to)} & & & & \\
        {\bf Benchmark} & {\bf Clauses} & {\bf Violations}
      & {\bf Positives} & {\bf AV} & {\bf AV} & {\bf SLOC}
      & {\bf Time (s)} & {\bf Static} & {\bf Final} \\
%        & & & {\bf \scriptsize (Conservative points-to)} & & & & \\
    \midrule
    %                                  cwords violations fposi pot  av  sloc    time   ICF static  ICF final
    Allocate Vector~\cite{IBM}        &  1  &   1      &  0   & 0  & 1 &  183 & 0.120 &      -   &     - \\
    Coord03~\cite{Artho03}            &  4  &   1      &  0   & 0  & 1 &  151 & 0.093 &      -   &     - \\
    Coord04~\cite{Artho04}            &  2  &   1      &  0   & 0  & 1 &   35 & 0.039 &      -   &     - \\
    Jigsaw~\cite{Praun2004}           &  1  &   1      &  0   & 0  & 1 &  100 & 0.044 &    121   &     2 \\
    Local~\cite{Artho03}              &  2  &   1      &  0   & 0  & 1 &   24 & 0.033 &      -   &     - \\
    Knight~\cite{Teixeira2011}        &  1  &   1      &  0   & 0  & 1 &  135 & 0.219 &      -   &     - \\
    NASA~\cite{Artho03}               &  1  &   1      &  0   & 0  & 1 &   89 & 0.035 &      -   &     - \\
    Store~\cite{PessanhaMSc}          &  1  &   1      &  0   & 0  & 1 &  621 & 0.090 &      -   &     - \\
    StringBuffer~\cite{Artho04}       &  1  &   1      &  0   & 0  & 1 &   27 & 0.032 &      -   &     - \\
    UnderReporting~\cite{Praun2004}   &  1  &   1      &  0   & 0  & 1 &   20 & 0.029 &      -   &     - \\
    VectorFail~\cite{PessanhaMSc}     &  2  &   1      &  0   & 0  & 1 &   70 & 0.048 &      -   &     - \\
    Account~\cite{Praun2004}          &  4  &   2      &  0   & 0  & 2 &   42 & 0.041 &      -   &     - \\
    Arithmetic DB~\cite{Teixeira2011} &  2  &   2      &  0   & 0  & 2 &  243 & 0.272 &      -   &     - \\
    Connection~\cite{Beckman2008}     &  2  &   2      &  0   & 0  & 2 &   74 & 0.058 &      -   &     - \\
    Elevator~\cite{Praun2004}         &  2  &   2      &  0   & 0  & 2 &  268 & 0.333 &      -   &     - \\
    \midrule
    %                                  cwords violations fposi pot  av  sloc time
    OpenJMS 0.7                       &  6  &   54     &  10  & 28 & 4 & 163K & 148   &     126  &    15 \\
    Tomcat 6.0                        &  9  &  157     &  16  & 47 & 3 & 239K & 3070  &     365  &    12 \\
    Cassandra 2.0                     &  1  &   60     &  24  & 15 & 2 & 192K & 246   &      -   &     - \\
    Derby 10.10                       &  1  &   19     &   5  &  7 & 1 & 793K & 522   &     122  &    16 \\
    Lucene 4.6                        &  3  &  136     &  21  & 76 & 0 & 478K & 151   &     391  &     2 \\
    \bottomrule
  \end{tabular}
  }
\end{table*}

% tomcat
%    158.828s 186.227s
%    162.837s 190.343s
%    160.433s 215.407s
%    159.747s 203.964s
%    158.920s 186.450s
%    159.747s 203.964s
%    162.753s 2889.829s
%    156.876s 245.983s
%    156.553s 185.037s
%
% t: 1436.694 4507.204
%
% lucene
%     76.723s 139.479s
%    159.747s 203.964s
%    159.747s 203.964s
%
% t:  396.217s 547.407s
%
% derby
%     90.170s 612.331s
%
% openjms
%    149.718s 168.453s
%    149.017s 185.791s
%    150.515s 169.618s
%    149.017s 185.791s
%    151.112s 168.655s
%    157.549s 177.100s
%
% t: 906.928s 1055.408s
%
% cassandra
%    139.105s 385.907s
%
% under            38.921-38.950s
% NASA             37.333-37.368s
% Local            38.579-38.612s
% StringBuffer     39.243-39.275s
% Jigsaw           38.990-39.034s
% Coord04          38.598-38.637s
% Knight           37.403-37.622s
% Coord03          38.877-38.971s
% Account          39.165-39.206s
% Allocate Vector  38.632-38.752s
% VectorFail       37.378-37.426s
% Store            39.616-39.706s
% Connection       38.669-38.727s
% Arithmetic DB    42.558-42.830s
% Elevator         40.038-40.371s

% explain table columns
Table~\ref{tab:valresults} summarizes the results that validate the
correctness of our approach.
The table contains both the macro benchmarks (above) and the micro benchmarks
(bellow).  The columns represent the number of clauses of the contract (Clauses);
the number of violations of those clauses (Contract Violations);
the number of false positives, i.e. sequences of calls that in fact the program
will never execute (False Positives);
the number of potential atomicity violations, i.e. atomicity violations that
could happen \emph{if} the object was concurrently accessed by multiple
threads (Potential AV); the number of real atomicity violations that can in
fact occur and compromise the correct execution of the program (Real AV);
the number lines of code of the benchmark (SLOC); and the time it took for the
analysis to complete (the analysis run time excludes the Soot initialization
time, which were always less than $179 \mathrm{s}$ per run).

% discuss results
Our tool was able to detect all violation of the contract by the client
program in the microbenchmarks, so no false negatives occurred, which supports
the soundness of the analysis.
Since some tests include additional contract clauses with call sequences
not present in the test programs we also show that, in general,
the analysis does not detect spurious violations, i.e.,
false positives.\footnote{In these tests no false positives were detected.
  However it is possible to create situations where false positives occur.
  For instance, the analysis assumes a loop may iterate an arbitrary number
  of times, which makes it consider execution traces that may not be possible.}
A corrected version of each test was also verified and the prototype correctly
detected that all contract's call sequences in the client program were
now atomically executed. Correcting a program is trivial since the prototype
pinpoints the methods that must be made atomic, and ensures the synchronization
required has the finest possible scope, since it is the method that corresponds to
the \emph{lowest} common ancestor of the terminals in the parse tree.

The large benchmarks show that gluon can be applied to large scale programs
with good results.
Even with a simple automated contract generation we were able to detect $10$
atomicity violations in real-world programs.
Six of these bugs where reported
(\href{https://issues.apache.org/bugzilla/show_bug.cgi?id=56784}{Tomcat}
\footnote{\url{https://issues.apache.org/bugzilla/show_bug.cgi?id=56784}},
\href{https://issues.apache.org/jira/browse/DERBY-6679}{Derby}
\footnote{\url{https://issues.apache.org/jira/browse/DERBY-6679}},
\href{https://issues.apache.org/jira/browse/CASSANDRA-7757}{Cassandra}
\footnote{\url{https://issues.apache.org/jira/browse/CASSANDRA-7757}}),
with two bugs already fixed in Tomcat 8.0.11.
The false positives incorrectly reported by gluon were all due to
conservative points-to information, since the program loads and calls
classes and methods dynamically (leading to an incomplete points-to graph).

ICFinder~\cite{liu2013} uses a static analysis to detect two types of
common incorrect composition patterns. This is then filtered with a dynamic
analysis.
Of the atomicity violations detected by gluon none of them was captured by
ICFinder, since they failed to match the definition of the patterns.

It is hard to directly compare both approaches since they use very different
approaches.  Loosely speaking, in Table~\ref{tab:valresults} the ICFinder Static
column corresponds to the Contract Violations, since they both represent the
static methodology of the approaches.  The ICFinder Final column cannot be
directly compared with the Real AV column because ``ICFinder Final'' may
contain scenarios that not represent atomicity violations (in particular if
ICFinder does not correctly identify the atomic sets).  ICFinder Final cannot
also be compared with ``Potencial AV'' since ``Potencial AV'' is manually
obtained from ``Contract Violations'', and ``ICFinder Final'' is the
dynamic filtration of ``ICFinder Static''.  In the end the number of bugs
reported by gluon was $6$, with $2$ bugs confirmed and with fixes already applied,
$1$ bug considered highly unlikely, and $3$ bugs pending confirmation;
ICFinder has $3$ confirmed and fixed bugs on
Tomcat.~\footnote{\url{https://issues.apache.org/bugzilla/show_bug.cgi?id=53498}}

% perf discuss results
The performance results show our tool can run efficiently.
For larger programs we have to use class scope mode, sacrificing precision
for performance, but we still can capture interesting contract violations.
The performance of the analysis depends greatly on the number of branches
the parser explores.
This high number of parsing branches is due to the complexity of the control flow of
the program, offering a huge amount of distinct control flow paths.
In general the parsing phase will dominate the time complexity of the analysis,
so the analysis run time will be proportional to the number of explored
parsing branches.
Memory usage is not a problem for the analysis, since the asymptotic space complexity
is determined by the size of the parsing table and the largest parsing tree.
Memory usage is not affected by the number of parsing trees because
our $GLR$ parser explores the parsing branches in-depth instead of in-breadth.
In-depth exploration is possible because we never have infinite height parsing
trees due to our detection of unproductive loops.

\subsection{ICFinder}
ICFinder tries to infer automatically what a module is, and incorrect compositions
of pairs of calls to modules.

Two patterns are used to detect potencial atomicity violations in method calls
compositions:

\begin{itemize}
\item USE: Detects stale value errors.  This pattern detects data or control flow
  dependencies between two calls to the module.
\item COMP: If a call to method \lstinline{a()} dominates \lstinline{b()} and
  \lstinline{b()} post-dominates \lstinline{a()} in some place, that is captured by
  this pattern.  This means that, for each piece of code involving two calls to the
  module (\lstinline{a()} and \lstinline{b()}), if \lstinline{a()} is always executed
  before \lstinline{b()} and \lstinline{b()} is always executed after \lstinline{a()},
  it is a COMP violation.
\end{itemize}

Both this patterns are extremely broad and contain many false positives.
To deal with this the authors filter this results with a dynamic analysis that
only consider violations as defined in~\cite{Vaziri2006}.
This analysis assumes that the notion of \emph{atomic set} was correctly inferred by
ICFinder.

\section{Related Work}
\label{sec:relwork}
%
% design by contract
The methodology of design by contract was introduced by Meyer~\cite{Meyer1992}
as a technique to write robust code, based on contracts between programs and
objects.
In this context, a contract specifies the necessary conditions the program must
met in order to call the object's methods, whose semantics is ensured if
those pre-conditions are met.

% design by contract for protocol checking
Cheon et al. proposes the use of contracts to specify protocols for accessing
objects~\cite{Cheon2007}.
These contracts use regular expressions to describe the sequences of calls
that can be executed for a given \emph{Java} object.
The authors present a dynamic analysis for the verification of the contracts.
This contrasts to our analysis which statically validates the contracts.
Beckman et al. introduce a methodology based on \emph{typestate} that statically
verifies if a protocol of an object is respected~\cite{Beckman2008}.
This approach requires the programmer to explicitly \emph{unpack} objects
before it can be used.
Hurlin~\cite{Hurlin2009} extends the work of Cheon to support protocols in concurrent
scenarios.
The protocol specification is extended with operators that allow methods to be
executed concurrently, and pre-conditions that have to be satisfied before the
execution of a method.
This analysis is statically verified by a theorem prover. Theorem proving,
in general, is very limited since automated theorem proving tend to be inefficient.

Peng Liu et al. developed a way to detect atomicity violations caused by
method composition~\cite{liu2013}, much like the ones we describe in this
paper.
They define two patterns that are likely to cause atomicity violations, one
capturing stale value errors and the other one by trying to infer a correlation
between method calls by analyzing the control flow graph
(if \lstinline{a()} is executed before \lstinline{b()}
and \lstinline{b()} is executed after \lstinline{a()}).
This patterns are captured statically and then filtered with a dynamic analysis.

Many works can be found about atomicity violations.
%Some of these methodologies can potentially be adapted to automatically infer
%the module's contract as defined in this paper.
Artho et al. in \cite{Artho03} define the notion of \emph{high-level data races},
that characterize sequences of atomic operations that should be executed
atomically to avoid atomicity violations.
The definition of high-level data races do not totally capture the
violations that may occur in a program.
Praun and Gross~\cite{Praun2004} extend Artho's approach to detect potential
anomalies in the execution of methods of an object and increase the precision of
the analysis by distinguish between read and write accesses to variables shared
between multiple threads.
An additional refinement to the notion of high-level data races was
introduced by Pessanha in~\cite{Pessanha2012}, relaxing the properties
defined by Artho, which results in a higher precision of the analysis.
Farchi et al.~\cite{Farchi2012} propose a methodology to detect atomicity
violations in the usage of modules based on the definition of high-level data races.
Another common type of atomicity violations that arise when sequencing several
atomic operations are \emph{stale value errors}.
This type of anomaly is characterized by the usage of values obtained atomically
across several atomic operations. These values can be outdated and compromise the
correct execution of the program. Various analysis were developed to detect these
types of anomalies~\cite{Artho04,Burrows2004,Pessanha2012}.
Several other analysis to verify atomicity violations can be found in the
literature, based on access patterns to shared
variables~\cite{Vaziri2006, Teixeira2011},
type systems~\cite{Caires2013}, semantic invariants~\cite{Demeyer2012},
and other specific methodologies~\cite{Flanagan2004,Flanagan2008,Flanagan2010}.

%\td{review. add paper %http://www.cs.ucla.edu/~lesani/companion/cav14/CAV14.pdf
%}

\section{Concluding Remarks}
\label{sec:conclusion}
In this paper we present the problem of atomicity violations when using a
module, even when their methods are individually synchronized by some
concurrency control mechanism.
We propose a solution based on the design by contract methodology.
Our contracts define which call sequences to a module should be executed in
an atomic manner.

We introduce a static analysis to verify these contracts. The proposed
analysis extracts the behavior of the client's program with respect to the module
usage, and verifies whether the contract is respected.

A prototype was implemented and the experimental results shows the analysis
is highly precise and can run efficiently on real-world programs.

% \listoftodos

\bibliographystyle{plain}
\bibliography{contracts,atomicity,misc}

\section{Appendix}
\subsection{Grammar Example}

\begin{figure}[H]
  \centering
  \begin{minipage}{0.95\linewidth}
    \begin{lstlisting}[frame=trbl]
@Contract(clauses=...)
class Module
{
    public Module() { }
    public void a() { }
    public void b() { }
    public void c() { }
}

public class Main
{
    private static Module m;

    private static void f()
    {
        m.c();
    }

    @Atomic
    private static void g()
    {
        m.a();
        m.b();
        f();
    }

    public static void main(String[] args)
    {
        m=new Module();

        for (int i=0; i < 10; i++)
            if (i%2 == 0)
                m.a();
            else
                m.b();

        f();
        g();
    }
}
    \end{lstlisting}
  \end{minipage}\hfill
  \caption{Program.}
%  \label{}
\end{figure}

\begin{figure}[H]
  \centering
  \begin{verbatim}
Start: A
A -> B
D -> E F
E -> X
F -> G
G -> H
B -> C
C -> D
L -> M
L -> N
M -> O
N -> R
O -> a P
H -> I
I -> K
I -> J
J -> L
K -> T U
BF -> BG
U -> V W
BG -> b BH
T -> BA
BH -> T BI
W -> epsilon
BI -> epsilon
V -> BD
BB -> c BC
Q -> S
BC -> epsilon
P -> Q
BD -> BE
S -> H
BE -> a BF
R -> b Q
Y -> Z
X -> Y
Z -> epsilon
BA -> BB
  \end{verbatim}
  \caption{Non-optimized grammar.}
%  \label{}
\end{figure}

\begin{figure}[H]
  \centering
  \begin{verbatim}
Start: A'
A' -> A
A -> I
L -> b I
L -> a I
I -> L
I -> T V
T -> c
V -> a b T
  \end{verbatim}
  \caption{Optimized grammar.}
%  \label{}
\end{figure}

\end{document}